\documentclass[runningheads]{llncs}
\usepackage{graphicx}
\usepackage{todonotes} 

\usepackage{comment}

\begin{document}
\title{Towards a Closer Collaboration Between Practice and Research in Agile Software Development Workshop: A Summary and Research Agenda}
\titlerunning{Towards a closer collaboration between practice and research}
%
\author{Michael Neumann\inst{1}\orcidID{0000-0002-4220-9641} \and
Eva-Maria Schön\inst{2}\orcidID{0000-0002-0410-9308} \and
Mali Senapathi\inst{3}\orcidID{0000-0003-3083-8069} \and
Maria Rauschenberger\inst{2}\orcidID{0000-0001-5722-576X} \and
Tiago Silva da Silva\inst{4}\orcidID{0000-0001-8459-7833} }
\authorrunning{M. Neumann et al.}
%
\institute{University of Applied Sciences Hannover, Hannover, Germany 
\email{michael.neumann@hs-hannover.de}\\
\and
University of Applied Sciences Emden/Leer,  Emden/Leer , Germany
\email{firstname.lastname@hs-emden-leer.de}\\
\and
Auckland University of Technology, Auckland, New Zealand\\
\email{mali.senapathi@aut.ac.nz}
\and
Federal University of S\~ao Paulo, S\~ao Paulo, Brazil\\
\email{silva.tiago@unifesp.br}
}
\maketitle              
\begin{abstract}
Agile software development principles and values have been widely adopted across various industries, influencing products and services globally. 
Despite its increasing popularity, a significant gap remains between research and practical implementation.
This paper presents the findings of the first international workshop designed to foster collaboration between research and practice in agile software development. 
We discuss the main themes and factors identified by the workshop participants that contribute to this gap, strategies to bridge it, and the challenges that require further research attention.

\end{abstract}
\section{Introduction}
Agile software development has gained high interest in research and practice for more than 20 years. Many phenomena have been examined in detail, and a considerable body of knowledge has been created. Examples of this include today's detailed understanding of human-centered software development~\cite{Kuchel.2023,Neumann.2024,Schoen.2023,Smite.2020}, agile requirements~\cite{Schoen.2017}, the scaling of agile approaches~\cite{Dingsoeyr.2019,Dingsoeyr.2018}, measuring agility~\cite{Looks.2021} and the well-known challenges posed by remote/distributed work~\cite{Neumann.2022a}. However, a significant gap persists between the research on agile software development and practice. The key factors that have led to this gap include the evolving, complex, and multifaceted nature of practitioner problems, the perception of research as irrelevant to practice, the complexity of organizational contexts, and the difficulty in addressing systemic issues~\cite{Barroca.2018}. A study highlights the importance of fostering collaboration and joint research efforts between academia and practice to bridge the gap between academic research and industry practices in managing requirements in agile software development~\cite{Dalpiaz.2024}. This collaboration can help raise awareness and address challenges and opportunities at the intersection of requirements and agile software development~\cite{Dalpiaz.2024,Gregory.2016,Wohlin.2012}. Other examples include integrating user experience (UX) into agile projects~\cite{Schoen.2023}, DevOps in agile~\cite{Senapathi2018}, large-scale agile, dependency management in agile projects, and applying agile values and practices in higher education~\cite{Schon2023a}.

To ensure that academic research addresses specific practitioner needs effectively, we identified two highly relevant gaps to start with: 
\begin{enumerate}
    \item The \textit{theory-gap} in agile research: Research within the agile community mainly consists of empirical case studies, typical in an interdisciplinary field like agile software development. However, practitioners express concerns about the external validity of case studies and question whether the findings can be generalized to the broader industry~\cite{Winters.2024}. Furthermore, many academic contributions reveal other aspects of the theory gap. A recent review of large-scale agile methods highlights several theoretical and practical issues in the existing literature, such as an overemphasis on the practices of agile frameworks at the expense of the foundational theoretical principles outlined in the Agile Manifesto~\cite{Beck.2001}. For instance, while large-scale agile transformations have been widely studied, there is a noticeable lack of theoretical developments on managing and sustaining their implementation~\cite{Carroll.2023}. Therefore, academic research is vital to develop theories that can enhance agile practices and processes and improve software outcomes, and thus provide a solid foundation for future empirical research~\cite{Schmid.2021,Winters.2024}.
    \item The \textit{time-gap} between practice and research: For two decades, agile practice has outpaced research. The adaptation of continuous improvement processes has resulted in a wide variety of agile practices in the industry. Furthermore, since 2020,  we have been experiencing an era of rapidly accelerating disruptive changes, such as the transition to remote work and the integration of Artificial Intelligence (AI) technologies into existing software development processes. In practice, while agile teams, particularly those with high maturity, demonstrate resilience in adapting to address these challenges, research is struggling to keep up and, metaphorically speaking, to get into the “\textit{driver's seat}” to actively shape these emerging trends. It is essential to balance the different timescales between research and industrial problems to produce research outcomes that are relevant and valuable to the industry~\cite{Woods.2025}.
\end{enumerate}

This paper summarizes the half-day workshop held at the International Conference on Agile Software Development, XP 2025 (XP2025). 
The workshop aimed to establish a foundation for fostering closer collaboration between researchers and practitioners, ensuring that academic research better addresses the needs of practitioners. 
It provided an interactive platform for sharing knowledge and exchanging experiences between academics and industry professionals.

\section{Workshop Results}
\label{sec:Results}
This section presents the results of our workshop, held on Monday, 2nd June 2025, as part of the XP2025 conference.
It consisted of three parts: (1) The first part was dedicated to setting the stage, clarifying the workshop's objectives, and ensuring a shared understanding among participants by introducing and explaining the definitions of the theory-gap and time-gap. (2) This was followed by a moderated panel discussion featuring four invited panelists, which lasted approximately 60 minutes. Participants were encouraged to contribute actively by asking questions and making comments. (3) The third part consisted of a \textit{World Café} session (approximately 60 minutes), during which participants engaged in group discussions on the identified gaps. The workshop ended with collaborative data synthesis. Further material on the workshop, such as slides and photos of the results, are available in a repository~\cite{Neumann.2025}.


\subsection{Panel Discussion}
In this part of the workshop, a panel with 4 panelists was held. The panelists included: Karen Eilers, Owner of the Institute of Information in Hamburg and a Lecturer at both the University of St. Gallen in Switzerland and Offenburg University of Applied Sciences; Nan Yang, Co-founder and Product Owner at Man Yi Education Consulting from 2017 to 2021, currently pursuing a PhD on ambidextrous software startups at LUT University in Lahti, Finland; Marcelo Luis Walter, Consulting Business Unit Manager at Objective Solutions in Brazil and a PhD candidate in Artificial Intelligence at the Pontifical Catholic University of Paraná in Curitiba, Brazil; and Tiago Silva da Silva, a Professor at the Federal University of São Paulo, Brazil. 

\subsection{World Café session}

In the World Café, workshop participants ($N=6$) expanded on their arguments through discussion and clustered them on brown paper using sticky notes. Photos of the activity are available in a repository~\cite{Neumann.2025}. The key points are summarized below.

\textit{The following key points regarding the \textbf{theory-gap} were discussed:}

\begin{itemize}
    \item Standardization of keywords.
    \item Abstracts without ``\textit{Click-bating description}''.
    \item Promote oral formats, such as videos, podcasts, and blog posts; researchers should also be evaluated based on these.
    \item Develop new KPIs beyond traditional papers' metrics; introduce alternative metrics for research evaluation (\textit{e.g.,} productivity) while addressing challenges in measuring different outcomes.
    \item Cross-disciplinary studies, action research, and longitudinal studies.
    \item Increase accessibility through more open-access research, but address the funding question: Who will cover the costs?
    \item Abstraction \textit{vs.} speed/return on investment.
    \item Researchers and practitioners working together on one project (risk of bias).
    \item Researchers should speak at practitioners' conferences and use simpler language.
\end{itemize}

\textit{The following key points regarding the \textbf{time-gap} were discussed:}

\begin{itemize}
    \item Split projects into more little pieces; overlapping phases.
    \item Maybe not always aim on closing the time-gap? \textit{``We need some time to make it right}''.
    \item Rapid release reporting, write papers and publish preprints; publish first, present later; medicine as a role model; this requires cultural shift.
    \item Propose small-scale experiments (MVPs).
    \item Grant models vs. incremental deliverables.
    \item Is the context chaotic or complex?
\end{itemize}

\section{Agenda for Future Research}
\label{sec:ResearchAgenda}
The industry’s primary focus remains on delivering impactful results and ensuring a strong return on investment~\cite{Winters.2024}. The workshop participants underscored the pivotal role that academia plays in validating the inherent unpredictability of the industry, where outcomes often stem from ill-informed decision-making, and reproducibility remains challenging due to a lack of clarity about how these results were achieved. This is where the contribution of academic research becomes vital: it seeks to clarify, through the scientific method, how successful strategies in one context can be adapted and applied to seemingly unrelated situations. As a result, any contributions from academic research that improve outcomes, reduce costs, improve efficiency, or facilitate the measurement of productivity are likely to attract significant interest~\cite{Winters.2024}. 

Several recommendations were proposed during the workshop to bridge the gaps between researchers and practitioners, particularly concerning theory and time. 
To address the theory-gap, suggestions included having researchers create diverse media formats — such as videos, podcasts, and blog posts — and using research methods, like Design Science Research and action research, that directly address industry challenges.
Additionally, making academic research more accessible through open access and presenting findings in simpler language for industry practitioners were highlighted as important steps. 

To mitigate the \textit{time-gap}, rapid dissemination of academic publications was suggested, including white papers and preprints. Central to addressing these gaps is the imperative to foster strong collaboration and meaningful relationships between researchers and practitioners.  
The challenges associated with \textit{time-} and theory-gap are apparent in \textit{Mode 1} research, which adheres to scientific principles but produces outcomes that are primarily relevant to the academic community, offering limited value to businesses~\cite{Huff.2001}. In contrast, \textit{Mode 2} research~\cite{MacLean.2002} adopts a collaborative model that bridges academic rigour with practical applicability by involving multi-stakeholder teams, comprising both academics and practitioners, who work together to address real-world problems~\cite{Gray.2011}. \textit{Mode 2} is built on several key principles: solving problems within specific application contexts, engaging in cross-disciplinary problem-solving that integrates theoretical and empirical elements, ensuring social accountability for research outcomes while cultivating a deeper understanding of diverse perspectives, and implementing quality controls that extend beyond traditional academic peer review to encompass the practical implications and impact of research~\cite{Barroca.2018}. This suggests that researchers must adopt a \textit{Mode 2} collaborative research model to address some of the challenges related to time and theory gaps. Nevertheless, \textit{Mode 2} comes (as every research model) with limitations. While applying \textit{Mode 2} the generalizability of the results can be one of the major threats to validity.  Although the workshop did not provide absolute predictions regarding future research problems or questions, the discussions indicated that numerous inquiries related to contemporary topics — such as the integration of AI and DevOps in agile development and large-scale agile transformations — could be explored using a \textit{Mode 2} collaborative approach to address the theoretical and temporal gaps discussed during the workshop.

\section{Conclusion}
\label{sec:Conclusion}
To address the challenges related to the \textit{theory-} and \textit{time- gap} between research and practice in agile software development, we organized the first international workshop at the XP2025 conference. This workshop provided an interactive platform for panelists and participants to critically reflect on and discuss the key issues in bridging the divide between researchers and practitioners in agile software development. Our aim was to identify the main themes related to theory and time gaps, foster a community of researchers interested in this subject, and develop an agenda for future research. We plan to continue building a community of researchers and practitioners, addressing the theory and time gaps through future workshops, and identifying specific industry-relevant research topics and questions that academia can explore.
%
%
%
 \bibliographystyle{splncs04}
 \bibliography{references}

\begin{thebibliography}{10}
\providecommand{\url}[1]{\texttt{#1}}
\providecommand{\urlprefix}{URL }
\providecommand{\doi}[1]{https://doi.org/#1}

\bibitem{Barroca.2018}
Barroca, L., Sharp, H., Salah, D., Taylor, K., Gregory, P.: Bridging the gap between research and agile practice: an evolutionary model. International Journal of System Assurance Engineering and Management  \textbf{9},  323--334 (2018). \doi{10.1007/s13198-015-0355-5}

\bibitem{Beck.2001}
Beck, K., Beedle, M., {van Bennekum}, A., Cockburn, A., Cunningham, W., Fowler, M., Greening, J., Highsmith, J., Hunt, A., Jeffries, R., Kern, J., Marick, B., Martin, R., Mellor, S., Schwaber, K., Sutherland, J., Thomas, D.: Agile manifesto (2021), \url{https://agilemanifesto.org/}

\bibitem{Carroll.2023}
Carroll, N., Conboy, K., Wang, X.: From transformation to normalisation: An exploratory study of a large-scale agile transformation. Journal of Information Technology  \textbf{38}(3),  267--303 (2023). \doi{10.1177/02683962231164428}

\bibitem{Dalpiaz.2024}
Dalpiaz, F., Steghöfer, J.P.: Where requirements and agility meet: No-man’s-land or a land of opportunity? IEEE Software  \textbf{41},  7--10 (2018). \doi{10.1007/s13198-015-0355-5}

\bibitem{Dingsoeyr.2019}
Dingsoeyr, T., Falessi, D., Power, K.: Agile development at scale: The next frontier. IEEE Software  \textbf{36}(2),  30--38 (2019). \doi{10.1109/MS.2018.2884884}

\bibitem{Dingsoeyr.2018}
Dingsøyr, T., Moe, N., Fægri, T., Seim, E.: Exploring software development at the very large-scale: a revelatory case study and research agenda for agile method adaptation. Empir Software Eng  \textbf{23}(2),  490--520 (2018). \doi{10.1007/s10664-017-9524-2}

\bibitem{Gray.2011}
Gray, D., Iles, P., Watson, S.: Spanning the hrd academic‐practitioner divide: bridging the gap through mode 2 research. Journal of European Industrial Training  \textbf{35},  247--263 (2011)

\bibitem{Gregory.2016}
Gregory, P., Barroca, L., Sharp, H., Deshpande, A., Taylor, K.: The challenges that challenge: Engaging with agile practitioners’ concerns. Information and Software Technology  \textbf{77},  92--104 (2016). \doi{https://doi.org/10.1016/j.infsof.2016.04.006}

\bibitem{Huff.2001}
Huff, A., Huff, J.: Re‐focusing the business school agenda. British journal of management  \textbf{12},  S49--S54 (2001)

\bibitem{Kuchel.2023}
Kuchel, T., Neumann, M., Diebold, P., Sch\"{o}n, E.M.: Which challenges do exist with agile culture in practice? In: Proc. of the Symposium on Applied Computing. p. 1018–1025 (2023). \doi{10.1145/3555776.3578726}

\bibitem{Looks.2021}
Looks, H., Fangmann, J., Thomaschewski, J., Escalona, M., Schön, E.M.: Towards a standardized questionnaire for measuring agility at team level. In: Agile Processes in Software Engineering and Extreme Programming. XP 2021. pp. 71–--85 (2021)

\bibitem{MacLean.2002}
MacLean, D., MacIntosh, R., Grant, S.: Mode 2 management research. British journal of management  \textbf{13},  189--207 (2002)

\bibitem{Neumann.2022a}
Neumann, M., Bogdanov, Y.: The impact of covid 19 on agile software development: A systematic literature review. In: Proceedings of the 55th Hawaii International Conference on System Sciences (2022). \doi{10.24251/HICSS.2022.882}

\bibitem{Neumann.2024}
Neumann, M., Kuchel, T., Diebold, P., Sch{\"o}n, E.M.: Agile culture clash: Unveiling challenges in cultivating an agile mindset in organizations. Computer Science and Information Systems  \textbf{21}(3),  1013--1031 (2024). \doi{10.2298/CSIS230715029N}

\bibitem{Neumann.2025}
Neumann, N., Schön, E.M., Senapathi, M., Rauschenberger, M., da~Silva, T.S.: Additional material to workshop agilepr at xp2025 (2025), \url{bit.ly/xp2025-agilepr1}

\bibitem{Schmid.2021}
Schmid, K.: If you want better empirical research, value your theory: On the importance of strong theories for progress in empirical software engineering research. In: Proceedings of the 25th International Conference on Evaluation and Assessment in Software Engineering. p. 359–364. Association for Computing Machinery, New York, NY, USA (2021). \doi{10.1145/3463274.3463360}

\bibitem{Schon2023a}
Sch{\"o}n, E.M., Buchem, I., Sostak, S., Rauschenberger, M.: Shift toward value-based learning: Applying agile approaches in higher education. In: Marchiori, M., Dom{\'i}nguez~Mayo, F.J., Filipe, J. (eds.) Web Information Systems and Technologies. pp. 24--41. Springer Nature Switzerland, Cham (2023)

\bibitem{Schoen.2017}
Schön, E., Winter, D., Escalona, M., Thomaschewski, J.: Key challenges in agile requirements engineering. In: Proceedings of the18th International Conference on Agile Software Development. Springer, Cham. (2017)

\bibitem{Schoen.2023}
Schön, E.M., {Silva da Silva}, T., Hinderks, A., Sharp, H., Thomaschewski, J.: Introduction to special issue on agile ux: challenges, successes and barriers to improvement. Information and Software Technology  \textbf{158},  107193 (2023)

\bibitem{Senapathi2018}
Senapathi, M., Buchan, J., Osman, H.: Devops capabilities, practices, and challenges: Insights from a case study. In: Proceedings of the 22nd International Conference on Evaluation and Assessment in Software Engineering 2018. p. 57–67. EASE '18, Association for Computing Machinery, New York, NY, USA (2018). \doi{10.1145/3210459.3210465}

\bibitem{Smite.2020}
{\v{S}}mite, D., Gonzalez-Huerta, J., Moe, N.B.: ``when in rome, do as the romans do'': Cultural barriers to being agile in distributed teams. In: Proc. of the Intl. Conf. on Agile Software Development. pp. 145--161 (2020)

\bibitem{Winters.2024}
Winters, T.: Thoughts on applicability. Journal of Systems and Software  \textbf{215},  112086 (2024). \doi{https://doi.org/10.1016/j.jss.2024.112086}

\bibitem{Wohlin.2012}
Wohlin, C., Aurum, A., Angelis, L., Phillips, L., Dittrich, Y., Gorschek, T., Grahn, H., Henningsson, K., Kagstrom, S., Low, G., Rovegard, P., Tomaszewski, P., van Toorn, C., Winter, J.: The success factors powering industry-academia collaboration. IEEE Software  \textbf{29}(2),  67--73 (2012). \doi{10.1109/MS.2011.92}

\bibitem{Woods.2025}
Woods, E.: Dear researchers step 1: Find a team with a problem. Journal of Systems and Software  \textbf{222},  112318 (2025)

\end{thebibliography}

\end{document}